\documentclass[useAMS,usenatbib]{mn2e}

\usepackage{graphicx}
\usepackage{rotating}
\usepackage[section] {placeins}
\usepackage{amsmath, amssymb}
\usepackage{color}
\usepackage{multirow}
\usepackage{makeidx}
\usepackage{xspace}
\usepackage{float}
\usepackage{subfig}

\newcommand{\degree}{$^{\circ}$}

\newcommand{\m}[1]{{\bf{#1}}}

\def\fp{\hbox{$.\!\!^{\reset@font\scriptscriptstyle\r@mn{p}}$}}

\long\def\crap#1{}

\def\log{\textrm{log}}

\def\<{\langle }
\def\>{\rangle }

\def\A3D{ATLAS$^\textrm{3D}$\xspace}

\def\A3D{ATLAS$^\textrm{3D}$\xspace}

\def\lS3{$\log\Sigma_{3}$}

\def\mdr{$T-\Sigma$\xspace}
\def\mmdr{$\langle$T$\rangle$--$\langle\Sigma\rangle$\xspace}
\def\kmdr{$kT-\Sigma$\xspace}
\def\m#1#2{\left[\begin{array}{c}
#1\\
#2\end{array}\right]}
\def\mm#1#2#3#4{\left[\begin{array}{cc}
#1 & #2 \\
#3 & #4 
\end{array}\right]}
\def\lodenscut{1.25\xspace}
\def\hidenscut{1.75\xspace}
\def\highEETGdenscut{1.5\xspace}
\def\nclustersindensbins{18, 30 and 8\xspace}
\def\nhighEfrac{7\xspace}
\def\nlowEfrac{26\xspace}

\def\fESig{f(E)--$\Sigma$\xspace}
\def\fSSig{f(S0)--$\Sigma$\xspace}
\def\fSISig{f(SI)--$\Sigma$\xspace}
\def\AvfESig{f($\langle$E$\rangle$)--$\langle\Sigma\rangle$\xspace}
\def\AvfSSig{f($\langle$S0$\rangle$)--$\langle\Sigma\rangle$\xspace}
\def\AvfSISig{f($\langle$SI$\rangle$)--$\langle\Sigma\rangle$\xspace}
\def\AvfEETGSig{f($\langle$E:ETG$\rangle$)--$\Sigma$\xspace}
\def\AvfSDISKSig{f($\langle$S0:Disk$\rangle$)--$\Sigma$\xspace}
\def\SRETG{$\langle$SR:ETG$\rangle$\xspace}
\def\EETG{$\langle$E:ETG$\rangle$\xspace}
\def\SDISK{$\langle$S0:Disk$\rangle$\xspace}
\def\sigsig{$\log\langle\Sigma_{3}\rangle$--$\log\Sigma_{\sigma}$\xspace}

\def\apld{$\log\langle\Sigma_{3}\rangle$\xspace}
\def\aagd{$\log\Sigma_\sigma$\xspace}

\title[Revisiting the original Morphology-Density Relation]{Revisiting the original Morphology-Density Relation}
\author[R. C. W. Houghton]{R. C. W. Houghton$^{1}$\thanks{Email: rcwh@astro.ox.ac.uk}\\
$^{1}$Physics Department, University of Oxford, Denys Wilkinson Building, Keble Road, Oxford, OX1 3RH, UK\\
}

\begin{document}

\date{}

\pagerange{\pageref{firstpage}--\pageref{lastpage}} \pubyear{2002}

\maketitle

\label{firstpage}

\begin{abstract}
In light of recent findings from the kinematic morphology-density relation, we investigate whether the same trends exist in the original morphology density relation, using the same data as \citeauthor{Dressler1980cat}. In addition to Dressler's canonical relations, we find that further refinements are possible when considering the \emph{average} local projected density of galaxies in a cluster.  
Firstly, the distribution of ellipticals in a cluster depends on the \emph{relative} local density of galaxies in that cluster: equivalent rises in the elliptical fraction occur at higher local densities for clusters with higher average local densities. This is not true for the late-type fraction, where the variation with local density within a cluster is independent of the average local density of galaxies in that cluster, and is as \citeauthor{Dressler1980} originally found.
Furthermore, the \emph{overall} ratio of ellipticals to early-types in a cluster does \emph{not} depend on the average density of galaxies in that cluster (unlike the ratio of lenticulars to disk systems), and is fixed at around 30\%. In the paradigm of fast and slow rotators, we show that such an elliptical fraction in the early-type population is consistent with a slow rotator fraction of 15\% in the early-type population, using the statistics of the ATLAS$^{3D}$ survey. 
We also find the scatter in the overall ratio of ellipticals to early-types is greatest for clusters with lower average densities, such that clusters with the \emph{highest} elliptical fractions have the \emph{lowest} average local densities. Finally, we show that average local projected density correlates well with global projected density, but the latter has difficulty in accurately characterising the density of irregular cluster morphologies.

\end{abstract}

\begin{keywords}

\end{keywords}

\section{Introduction}
\label{sec:intro}
It is well known that early-type galaxies (ETGs) are more common in denser (cluster) environments, with a corresponding decrease in numbers of late-type galaxies \citep[LTGs:][]{Oemler74,DavisGeller76}. \citet[][hereafter D80]{Dressler1980} convincingly demonstrated this morphology-density relation in a quantitative manner, showing how the number fraction of lenticulars (S0s) and, at higher densities ellipticals (Es), rises with local projected density ($\log\Sigma$) to replace a diminishing LTG (or spirals and irregulars, S+I) fraction. With morphology described by the \emph{T}-type, this relation was coined the \mdr relation. \citet{Whitmore1993} later showed that an equivalent relation holds with \emph{scaled} radius (the $T$--$R$ relation). \citet{Dressler1997} also extended the study to higher redshifts and found a lack of S0s and an excess of Es, suggesting very different formation epochs for the two morphologies.

Recently, integral-field spectroscopy (IFS) surveys such as SAURON \citep{SAURONI} and \A3D \citep{ATLAS3DI} have been able to refine the S0/E division, based on the internal kinematic properties of the galaxies. In the complete volume limited \A3D sample, two thirds of the Es show rapid rotation in their velocity maps, which are indistinguishable from the velocity maps of typical S0s. This led to the classification of fast rotators (FRs), a population consistent with being oblate axisymmetric spheroids \citep{SAURONX,ATLAS3DIII}. The remaining Es (and some S0s) showed an absence of rotation in their velocity maps leading to the classification of slow rotators (SRs), a population which may be (mildly) triaxial \citep{ATLAS3DIII}. 

The \mdr relation was revised by \citet{ATLAS3DVII}, who used the kinematic classifications FR/SR and the morphological LTG classification to develop a \emph{kinematic morphology-density relation} (\kmdr). This \kmdr relation shows a clear decline in the LTG number fraction towards higher projected density 
mirrored by a corresponding rise in the number density of FRs 
This is qualitatively similar to the previous results of D80 for the S0 and LTG morphological classes. \citet{ATLAS3DVII} also reported that the total fraction of SRs was constant at 4\% across all environments, except for the core of the Virgo cluster, where it rose to 20\%. This behaviour is noticeably different to that of Es in the original \mdr relation, which show a more gradual and sustained increase in numbers with projected density with the fraction of ellipticals reaching 40\%. 

However, the \A3D survey mostly sampled field and group galaxies, and contained only a single cluster (Virgo) which is unrelaxed, spiral rich, and $\times$100 lower in density than the nearby Coma cluster. Conversely, D80 studied 55 rich clusters and a single `field sample' to construct the original \mdr relation. Thus, the \A3D \kmdr and original \mdr relations probe very different types of environment and are not directly comparable. Motivated by this and the over-abundance of SRs in the core of Virgo, \citet{DEugenio2013} and \citet{Houghton2013} studied the \kmdr relation in two more massive, denser clusters: Abell~1689 and Coma. These studies highlighted that, while the internal fraction of ETGs and LTGs may vary, the \emph{average} fraction of SRs in the ETG population (the \SRETG fraction) is constant at around 15\%. This result was confirmed in lower mass and unrelaxed clusters \citep{Scott2014,Fogarty2014} and appears to hold across five orders of magnitude in projected density, from the field (\A3D data) to the densest cluster (Abell~1689). However, within each cluster, the SR:ETG ratio varies strongly with \lS3: more SRs were found in the densest (\emph{viz.} central) regions, but only at the expense of the less dense (\emph{viz.} outer) regions, where FRs were abundant. This led to the conclusion that the SRs were being segregated from the rest of the population, possibly via dynamical friction due to the massive nature of most SRs. Crucially, the SR:ETG ratio peaks at progressively higher \lS3 for clusters where the \emph{average} projected local density of galaxies is higher \citep{Houghton2013}. Indeed, at the highest projected galaxy density of Virgo (containing the highest SR fraction in Virgo), \citet{DEugenio2013} found no SRs in Abell~1689.

The implications of these finding are significant. The different kinematic structures of FRs and SRs are believed to originate from different assembly mechanisms, meaning the kinematic classifications directly relate to different assembly histories. One would not expect the efficiency of these two mechanisms to vary in the same way. It is often suggested that S0s result from quenched spiral galaxies while ellipticals originate via mergers \citep{ToomreToomre1972,BarnesHernquist1992,NaabTrujillo2006,ATLAS3DVI,ATLAS3DVIII}; the rapid rotation in FRs is strongly suggestive of a disk-like origin whereas the random, often decoupled, motions of a SR are indicative of a complicated, perhaps violent, assembly from multiple components. It is intriguing therefore that the overall density of galaxies in any environment feeds these two processes equally, producing a fixed ratio of FRs and SRs. Is the formation of SRs dependent on the availability of FRs (i.e. a dry-merger scenario)? Could FRs and SRs in fact be produced by a single mechanism (i.e. degrees of harassment or minor merging)? 

In this paper we investigate if recent results from the \kmdr relation also hold when using just visual morphologies. Specifically we aim to find out: how the E:ETG ratio varies with \lS3 within Virgo-like, Coma-like and Abell 1689-like clusters; and how the \EETG ratio of an entire cluster varies with the average local density of galaxies in that cluster. The structure of this paper is as follows: \S\ref{sec:sample} describes the samples from which the data was compiled, \S\ref{sec:results} describes the results, which are then discussed in \S\ref{sec:discussion}, and \S\ref{sec:conclusion} concludes. We assume standard WMAP7 cosmology throughout \citep{WMAP7COSMO}. 

\section{Samples \& Data}
\label{sec:sample}
The data used in the original \mdr was published in \citet{Dressler1980cat}. Using that catalogue, we updated the cluster redshifts using the latest values from NED\footnote{http://ned.ipac.caltech.edu} as well as the distances to the clusters, assuming all clusters are at rest with respect to the Hubble flow. We further checked the cluster redshifts using the galaxy redshifts available in NED \citep[mostly from the 6dF survey, ][]{6dF}, the 2MASS redshift survey \citep{TMZ}, the WINGS survey \citep{WINGSdata} and the CAIRNS survey \citep{Rines2003}. We found that the redshifts published by NED for the clusters Abell~838, Abell~1631 and DC2349-28 were inconsistent with the redshifts of all the galaxies: these three objects appear to be projections of at least two clusters along the line of sight. Further details are included in Appendix A; here we simply exclude these clusters from all subsequent analysis. \citet{DresslerSchectman1988} also found Abell~1631 to be a projection of multiple clusters, as well as Abell~1736; we do not reject Abell~1736 from our sample as the two components are only separated by 3500 km/s, which at face value only equates to a distance error of $\sim$15\% for each component. This results in a catalogue of 5455 galaxies in 53 clusters.\footnote{We do not combine data for DC 0107-46 and DC 0103-47, although considered a single cluster in D80.}

We generalise the more detailed morphological classes available in the \citet{Dressler1980cat} catalogue into LTGs (spirals and irregulars, S+I), lenticulars (S0) and ellipticals (E). 
\begin{figure}
   \centering
   \includegraphics[width=0.5\textwidth]{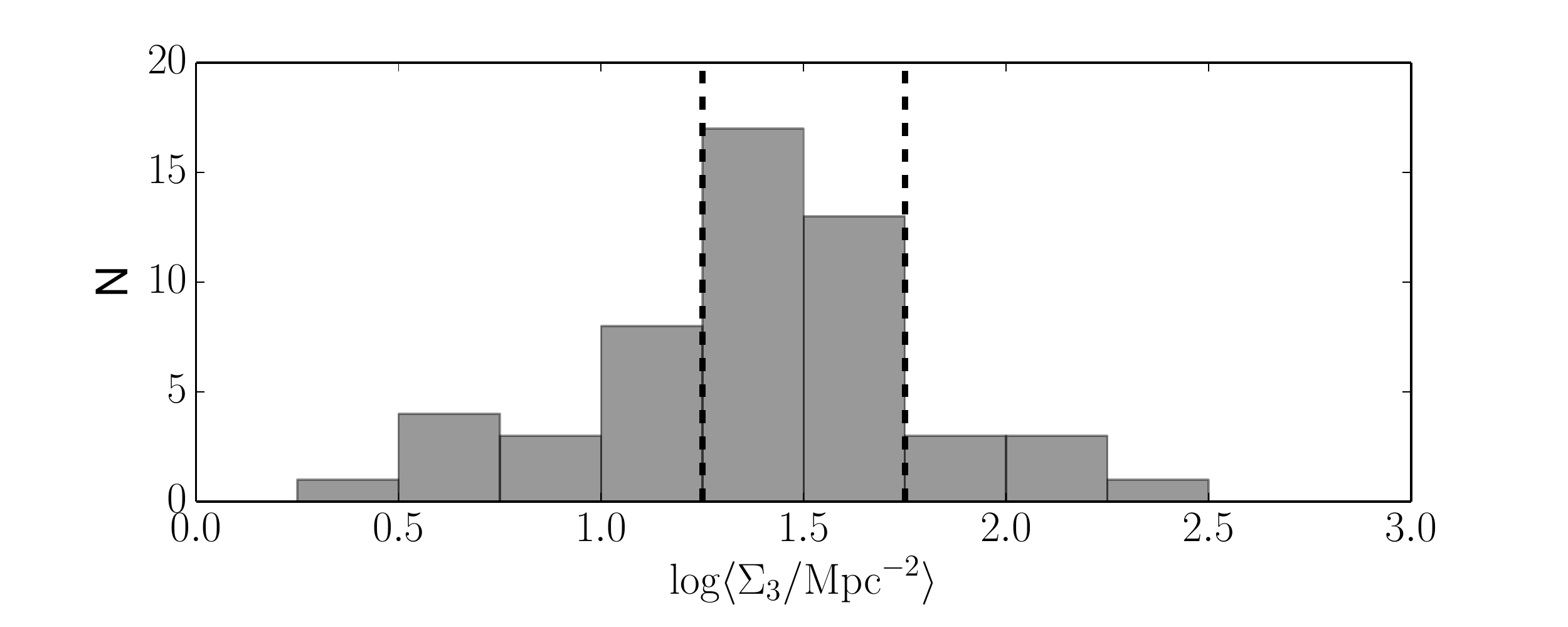} 
   \caption{The distribution of the average projected density of galaxies in each cluster, $\log\langle\Sigma_{3}/\textrm{Mpc}^{2}\rangle$. Dashed lines divide the sample into low-, medium- and high-density clusters. }
   \label{fig:mLogSig}
\end{figure}

\begin{figure}
   \centering
   \includegraphics[width=0.5\textwidth]{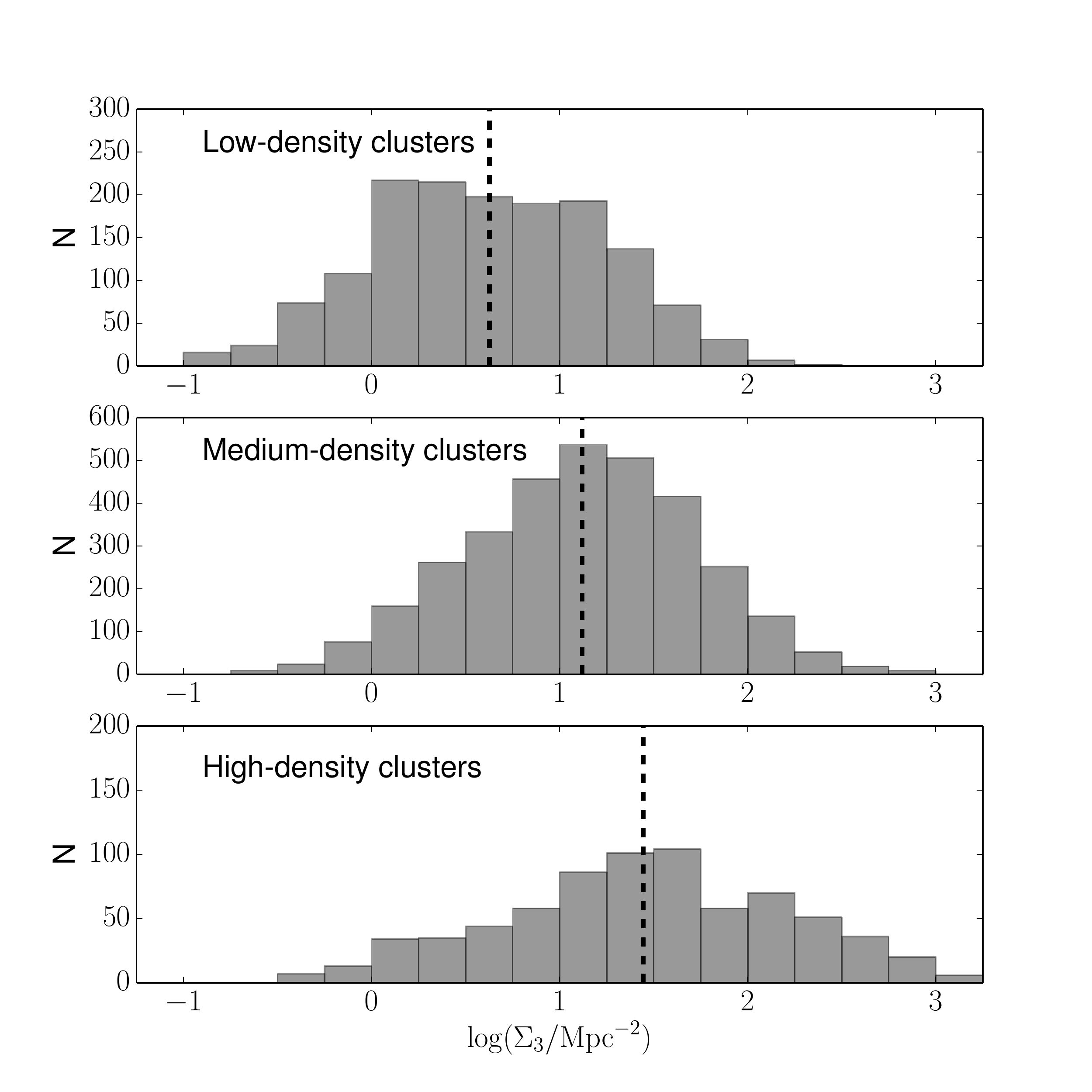} 
   \caption{The distribution of local projected galaxy density (\lS3) within low-, medium- and high-density clusters. The distributions are log-normal and don't exhibit particularly strong tails to low or high density. Thus the mean projected density of galaxies in each cluster is a fair representation of the typical projected density of galaxies in each cluster. }
   \label{fig:LogSigDists}
\end{figure}
The local projected densities of the galaxies were not published in \citet{Dressler1980cat} so we re-calculate them. We define the $N$th average projected density $\Sigma_{N}$ to be $N$ times the reciprocal of the smallest circular area ($a_{N}$ measured in square arc seconds, or $A_{N}$ when measured in kpc$^{2}$) at the distance of each cluster that encloses the nucleus of the $N$th nearest neighbour. Where galaxies are found at the edges of the photographic plates, we correct $a_{N}$ for the region missing of the plate edge. 
We adopt the same correction for foreground/background galaxies as D80 did, but instead of subtracting this correction, we incorporate it into the calculation of $a_{N}$: we count the number of galaxies \emph{above} what is expected by the background while increasing the circular area on the sky until the required number $N$ is attained. In one case, this approach fails as the number of galaxies never exceeds the background rate; we remove this galaxy from all subsequent analysis. In three cases, the number of galaxies above the background rate never reaches $N$, so we calculate $\Sigma_{N}$ using the peak value of the number of galaxies found above the background rate. For all the other 
5451 galaxies, this approach succeeds and provides a better measure of $\Sigma_{N}$ than subtracting a correction, \emph{a posteriori}: $N$ refers to the number of galaxies \emph{above} the background (not including it) which helps produce a Normal distribution, particularly at lower densities (c.f. Fig. \ref{fig:mLogSig} \& \ref{fig:LogSigDists}). 

We separate clusters into groups of different (projected) density by taking the average projected density of the galaxies in each cluster and defining low-, medium- and high-density clusters as those with average projected densities of $\log\langle\Sigma_{3}/\textrm{Mpc}^{-2}\rangle$$\leq$\lodenscut, \lodenscut$<$$\log\langle\Sigma_{3}/\textrm{Mpc}^{-2}\rangle$$\leq$\hidenscut and $\log\langle\Sigma_{3}/\textrm{Mpc}^{-2}\rangle$$>$\hidenscut. The number of clusters in each of these ranges is \nclustersindensbins, respectively. Fig. \ref{fig:mLogSig} shows the distribution of $\log\langle\Sigma_{3}\rangle$ and these divisions. In Fig. \ref{fig:LogSigDists} we show the distribution of $\log\Sigma_{3}$ for galaxies from clusters in each group. These distributions of $\log\Sigma_{3}$ are Normal-like and not obviously skewed towards higher or lower density for any density class. 
Hereafter, $\log\Sigma_{3}$ is used to refer to the projected local density of an individual galaxy, while $\log\langle\Sigma_{3}\rangle$ is used to refer to the mean projected local density of all galaxies in a cluster.

We calculate the \emph{average} E, S0 and S+I fractions in each cluster to compare with $\log\langle\Sigma_{3}\rangle$. At this point, we are able to correct for \emph{differential} foreground/background contamination using the fractions found in \citet{Dressler1997} and the areas of the photographic plates. In one case, the expected number of E contaminants is greater than the observed number of E galaxies; we therefore set the corrected number of Es to zero.

In order to better understand the role of $\log\langle\Sigma_{3}\rangle$, we compare it to a measure of \emph{global} density. We determine the median RA and Dec of all galaxies in each cluster and adopt this as the cluster centre. While we investigated luminosity weighted and mean centres, we found the median to be the most robust. We then calculated the standard deviation of the galaxy locations along the RA and Dec axes and took the geometrical mean as a fiducial measure of the size of the cluster (hereafter $\sigma_{XY}$). The sizes measured this way were always less than 1 Mpc. Counting the number of galaxies within a circle of radius $\sigma_{XY}$ centred on the median coordinate, we measured the average global density of galaxies $\Sigma_{\sigma}$ = N(r$<$$\sigma_{XY}$) / $(\pi\sigma_{XY}^{2})$. While the results of this density metric are qualitatively similar to an average within a fixed radius aperture (e.g. 1 Mpc), an adaptive region over which to calculate the density takes account of the relative differences in cluster sizes and morphologies; that said, there are three clusters for which even an adaptive circular aperture fails to encompass the bulk of the galaxies due to their highly non-spherical distribution of galaxies. This is discussed in more detail in \S\ref{sec:results}. 

For morphological fractions, we calculate \emph{posterior} values and uncertainties as in \citet{Houghton2013}; the difference between measured number fraction and posterior number fraction is usually small unless the fraction is close to zero or one, or if the total number of galaxies is small. Our uncertainties (and error bars) mark the 15.9 and 84.1 percentiles of the posterior probability distribution (i.e. $\pm1\sigma$); similarly, our reported values (symbols on plots) mark the median (50th percentile) of the posterior probability distribution. Shaded regions on plots are drawn between the 15.9 and 84.1 percentiles (i.e. they are equivalent to $\pm1\sigma$ error bars).

\section{Results}
\label{sec:results}

\begin{table}
\caption{Polynomial coefficients for the fits made to various relations between morphological type and projected density ($f=\sum_{i}c_{i} (\log\Sigma_{3})^{i}$). Relations 1-3 are shown in Fig. \ref{fig:D80_MDR}; relations 4-6 are shown in Fig. \ref{fig:AvMDR} and relation 7 is shown in Fig. \ref{fig:AvEETGfrac}. Uncertainties are only given for the (inverse variance weighted) linear fits.}

\begin{tabular}{lr@{$\pm$}lr@{$\pm$}lcc}
\textbf{Relation} &  \multicolumn{2}{|c|}{$\bf c_{0}$} &  \multicolumn{2}{|c|}{$\bf c_{1}$} & $\bf c_{2}$ & $\bf c_{3}$ \\
\hline
1: \fESig                 & \multicolumn{2}{|c|}{0.00579} & \multicolumn{2}{|c|}{0.02545} & 0.01570 & 0.11630 \\
2: \fSSig                 & \multicolumn{2}{|c|}{-0.00952} & \multicolumn{2}{|c|}{0.01434} & 0.08544 & 0.32620 \\
3: \fSISig                & \multicolumn{2}{|c|}{0.00462} & \multicolumn{2}{|c|}{-0.04104} & -0.10229 & 0.55909 \\
4: \AvfESig             & 0.065 & 0.02 & 0.063 & 0.01 & - & - \\
5: \AvfSSig             & 0.272 & 0.03 & 0.116 & 0.02 & - & - \\
6: \AvfSISig            & 0.601 & 0.03 & -0.178 & 0.02 & - & - \\
7: \AvfEETGSig       & 0.242 & 0.03 &  0.026  &  0.02 & - & - \\
8: \AvfSDISKSig       & 0.295 & 0.03 & 0.185 & 0.02 & - & - \\
9: \sigsig                 & \multicolumn{2}{|c|}{-0.249} &  \multicolumn{2}{|c|}{1.148} & - & - \\
\end{tabular}
\label{tab:fits}
\end{table}

We now present the original \mdr relation, the \mdr relations for clusters of different average densities, followed by the \mmdr and \EETG-$\langle\Sigma\rangle$ relations to study how the average morphological mix in a cluster depends on the average density of galaxies in that cluster. We further show how average local density relates to global density via the \sigsig relation. 

Where polynomial fits are made and shown in figures, the coefficients are given in Table \ref{tab:fits}. Linear least-squares fits were inverse-variance weighted for all except the \sigsig relation; for this we used a Bisector estimator in the absence of uncertainties on either axis \citep{Isobe1990}.  We tabulate the coefficients and uncertainties in Table \ref{tab:fits} and show 1$\sigma$ ranges for the fits as dotted lines in Figs. \ref{fig:AvMDR} \& \ref{fig:AvEETGfrac}. 

\begin{figure}
   \includegraphics[width=0.5\textwidth]{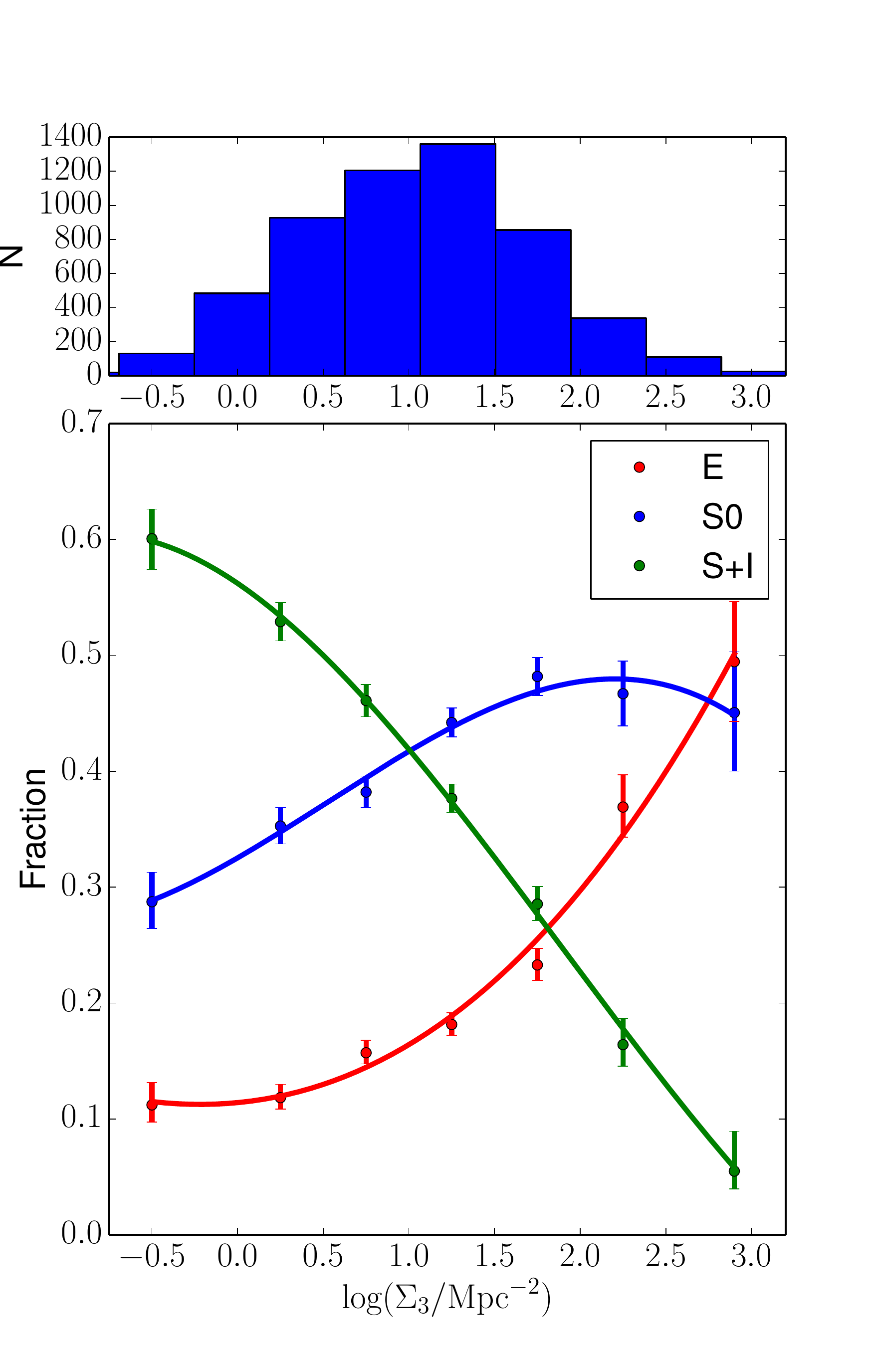} 
   \caption{Reproduction of the \mdr relation of D80 using up-to-date cluster redshifts, projected densities to the third nearest neighbour and field-edge or completeness corrections.}
   \label{fig:D80_MDR}
\end{figure}

\begin{figure}{\phantom{H}}
   \centering
   \includegraphics[width=0.5\textwidth]{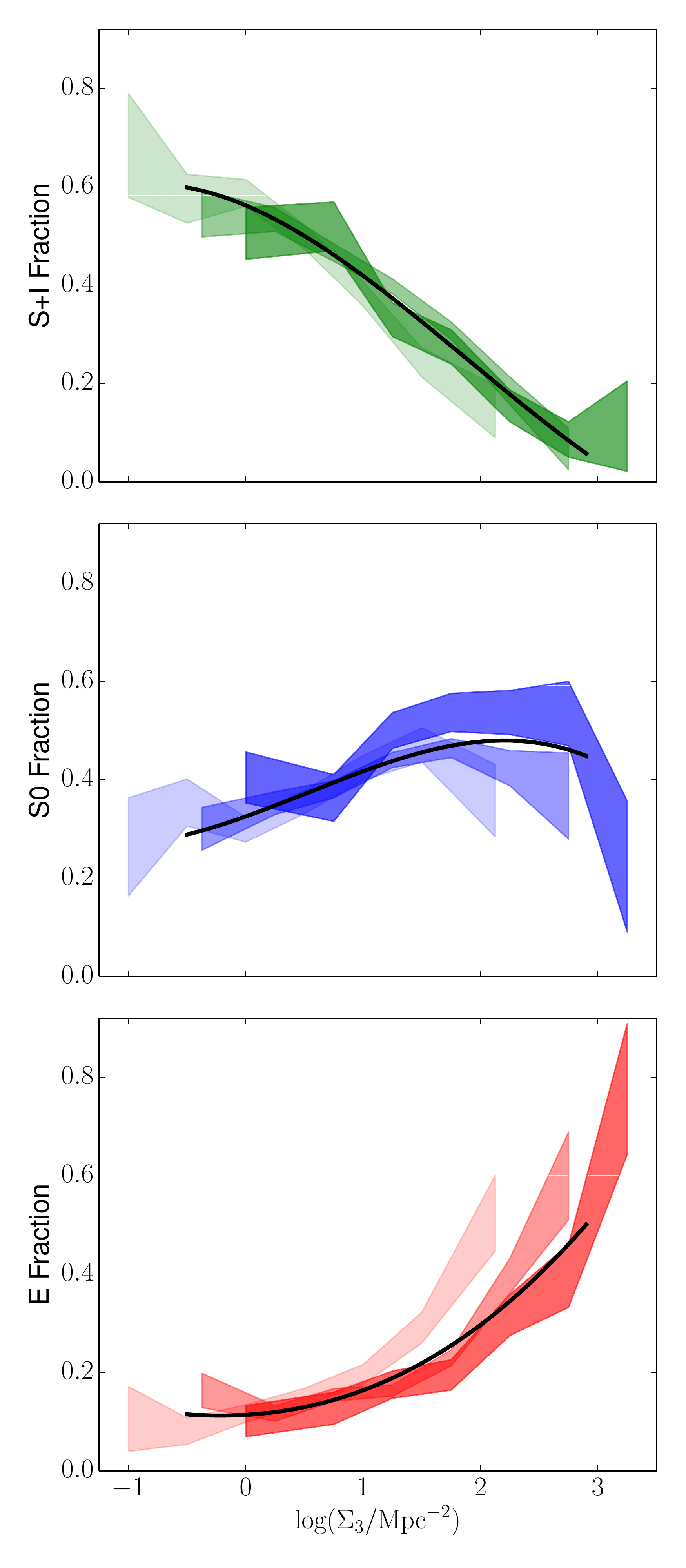} 
   \caption{The \mdr relations for different density clusters. In all panels, we show number fraction against local projected density for the low-, medium- and high-density clusters (light, medium and dark shading; defined in Fig. \ref{fig:mLogSig}); the black lines are the same as those shown in Fig. \ref{fig:D80_MDR}. \textbf{Top}: the S+I (LTG) fraction. \textbf{Middle}: the S0 fraction. \textbf{Bottom}: the E fraction. While the variation in the S+I and S0 number fractions is similar for clusters of different average density (and to the relations in Fig. \ref{fig:D80_MDR}), the E fraction is dependent on both the local projected density \emph{and} the average projected density galaxies in the cluster.}
   \label{fig:LMHMDR}
\end{figure}

\subsection{The original \mdr relation}

Fig. \ref{fig:D80_MDR} shows our re-processed data in the same style as presented by D80. Using the number fractions of LTGs, S0s and Es, we recover the same trends. We fit polynomials to the data (unweighted by uncertainties) to act as references for future comparisons (coefficients in Table \ref{tab:fits}). Note that our use of $\log\Sigma_{3}$ (rather than $\log\Sigma_{10}$ as in D80) shifts the trends to higher values of local density. Furthermore, the updating of the sample and the removal of three clusters results in a higher peak E-fraction. However, the value of this peak is dependent on the binning: to sample a larger number of galaxies and obtain better statistics, the bin at highest densities covers a larger range in local density (2.5$<\log\Sigma_{3}\leqslant3.3$). If we split this last bin into two, the E-fraction peaks at a higher value, but with considerably larger uncertainties; in the extreme case, we could define a bin that samples \emph{just} galaxies in the densest environments of the high-density clusters, thereby recovering the peak E-fraction that we will see in Fig. \ref{fig:LMHMDR}.

\subsection{The \mdr relation for low, medium and high-density clusters}

Fig. \ref{fig:LMHMDR} illustrates the \mdr relations for low-, medium- and high-density clusters. Accordingly, we introduce the following terms: the \fSISig relation describes the change in the number fraction of S+Is with \lS3; the \fSSig relation describes the change in the number fraction of S0s with \lS3 and the \fESig relation describes the change in the number fraction of Es with \lS3. 

The maximum (\emph{viz.} central) degree of segregation (i.e. the maximum E-fraction, maximum S0 fraction and minimum S+I fraction) appears similar in low-, medium- and high-density clusters. This is contrary to what one might expect: clusters with galaxy populations that extend to high \lS3 appear to have similar morphological fractions in their densest regions as clusters whose galaxy populations only extend to lower \lS3. The S0 fraction also appears to drop at the very densest regions; this was also seen by \citet{Whitmore1993}

Fig. \ref{fig:LMHMDR} also shows the best-fit polynomials from the \mdr relation in Fig. \ref{fig:D80_MDR} (over plotted as black lines). The variation of the \fSISig and \fSSig relations is broadly similar to these relations and all are close to linear in $\log\Sigma_{3}$. But the \fESig relation is quite different: it reaches a peak value of 50\%--70\%, irrespective of the average density of galaxies in the clusters. Furthermore, the shape of the \fESig relation is similar for low- medium- and high-density clusters, but shifted to higher densities for denser clusters. Note that the densities which have the highest fraction of Es in low-density clusters (or medium-density clusters) have a \emph{significantly} lower E fraction in high-density clusters. 

\subsection{The \mmdr relation}

\begin{figure} 
   \centering
   \includegraphics[width=0.5\textwidth]{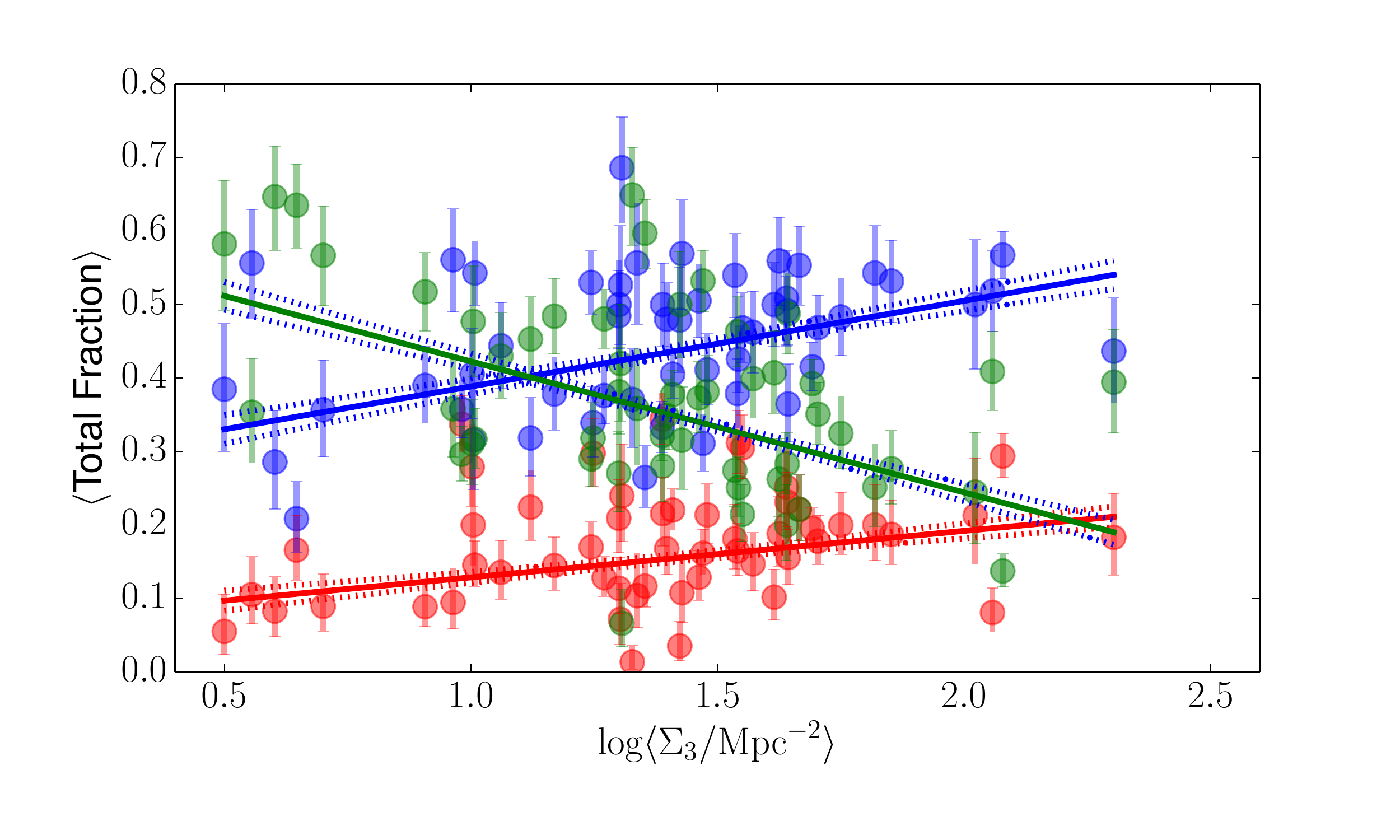} 
   \caption{The \mmdr relation, relating the average morphological fraction to the average projected density (for individual clusters). The variation with average density is weak and comparable to the intrinsic scatter in each relation.}
   \label{fig:AvMDR}
\end{figure}

Fig. \ref{fig:AvMDR} shows the morphological fractions and local galaxy densities averaged within each cluster; individual points represent different clusters while the lines show linear least-squared fits. The \mmdr relations are much weaker than the \mdr relations, varying only slightly with the $\log\langle\Sigma_{3}\rangle$ of each cluster. We measure the observed scatter $\sigma_{o}$ (and estimate the intrinsic scatter $\sigma_{i}$) in each relation to be 0.07 (0.06), 0.09 (0.07) and 0.11 (0.98) for the E, S0 and S+I fractions, respectively. The similarity between $\sigma_{o}$ and $\sigma_{i}$ suggests $\sigma_{i}$ is considerably greater than the measurement uncertainty. 
The scope of each linear relation ($\Delta f$) over the range in density probed ($\Delta \log\Sigma_{3} = 2.5 - 0.5$) is only 1.1, 1.7 and 1.8 (for E, S0 and S+Is, respectively) times the range spanned by the intrinsic scatter ($\pm1\sigma_{i}$). As such, we confirm that these relations are \emph{weak} in their predictive power. 

\begin{figure}
   \centering
   \includegraphics[width=0.5\textwidth]{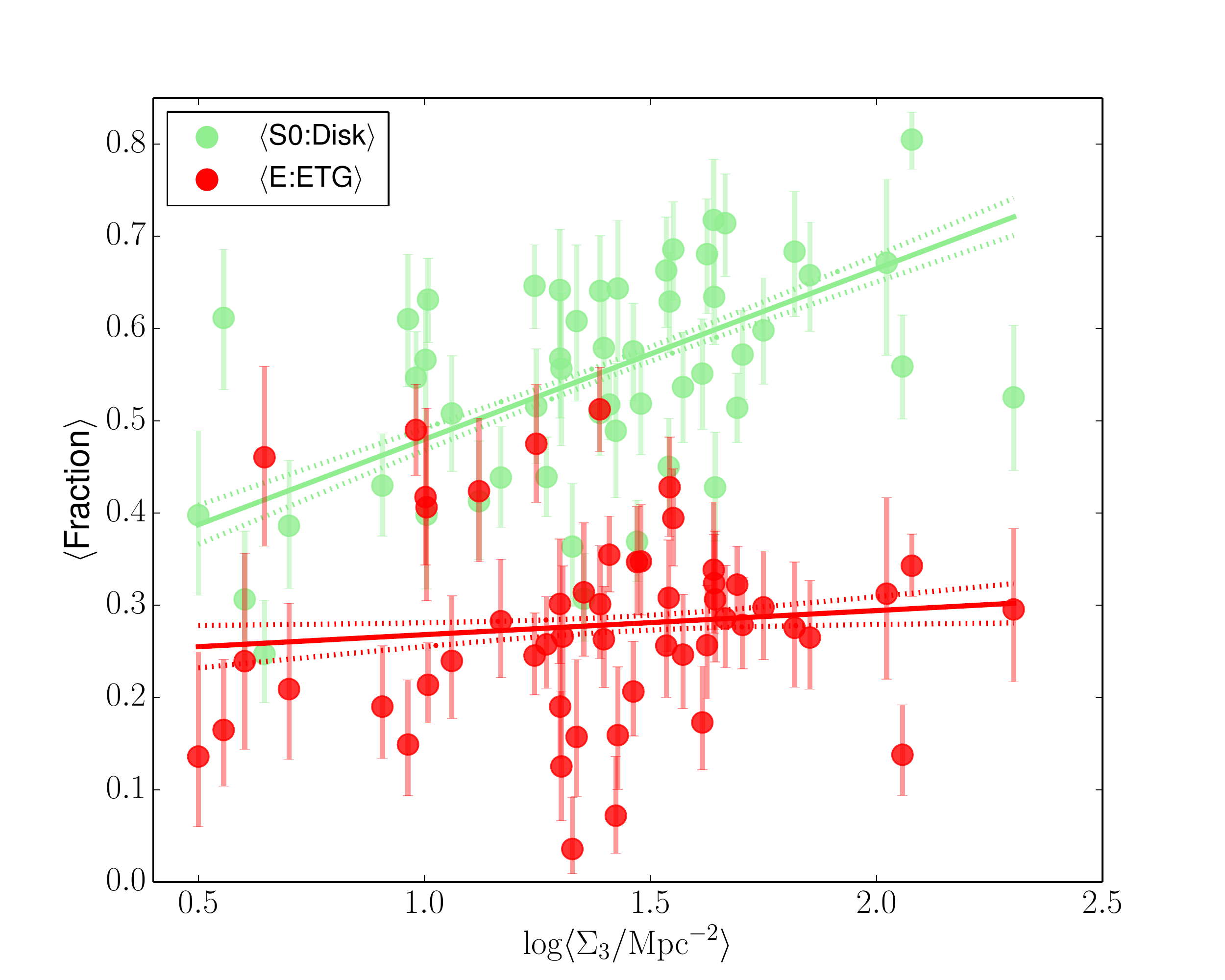}   
   \caption{The \EETG and \SDISK ratios (red and green points, respectively) averaged over individual clusters, versus $\log\langle\Sigma_{3}\rangle$. While the fraction of S0s in the disk population shows a steady increase with $\log\langle\Sigma_{3}\rangle$, the fraction of ellipticals in the ETG population is constant at around 30\% for all clusters.}
   \label{fig:AvEETGfrac}
\end{figure}

Fig. \ref{fig:AvEETGfrac} shows the mean S0 fraction \emph{within the disk population} (\SDISK = N[S0]/\{N[S0]+N[S+I]\}) and the mean E fraction \emph{within the ETG population} (\EETG=N[E]/\{N(E)+N(S0)\}) in each cluster as a function of $\log\langle\Sigma_{3}\rangle$. The \AvfSDISKSig relation shows an increase in the fraction of passive disk systems with $\log\langle\Sigma_{3}\rangle$. This is to be expected given Fig. \ref{fig:AvMDR}. Conversely, the \AvfEETGSig relation is consistent with a slope of zero and the \EETG ratio appears constant at around 30\%. However, there is significant scatter at lower densities ($\langle\log(\Sigma_{3}/\textrm{Mpc}^{2})\rangle < 1.5$) with \EETG ratios within individual clusters varying between 10\% and 50\%. If we estimate the mean \EETG ratio for low-, medium- and high-density clusters, we find 
0.296$\pm$0.03, 
0.271$\pm$0.02 and 
0.275$\pm$0.03, 
respectively (quoting the standard error in the mean). We note that the standard error in the mean does not take account of measurement errors.

Below $\log(\Sigma_{3}/\textrm{Mpc}^{-2})$=\highEETGdenscut, there are a group of clusters with remarkably high \EETG ratios. We can split these lower density clusters into those that have \EETG ratios above and below 40\%. The clusters in both divisions have similar distances (the mean redshifts are 0.051 and 0.043 for the high and low \EETG clusters, respectively) suggesting the difference is not a result of a distance-dependent misclassification error. Of the seven clusters with high \EETG ratios, five have Bautz-Morgan classifications in {\sc NED}\footnote{http://ned.ipac.caltech.edu}: they span all types (I, I/II, II/III and 2$\times$III). All but one of the clusters with low \EETG ratios have Bautz-Morgan classifications in {\sc NED}: again they span all types (6$\times$I, 2$\times$I/II, 5$\times$II, 2$\times$II/III and 10$\times$III). Thus the high and low \EETG clusters do not appear to have substantially different morphologies. 

\begin{figure} 
   \centering
   \includegraphics[width=0.5\textwidth]{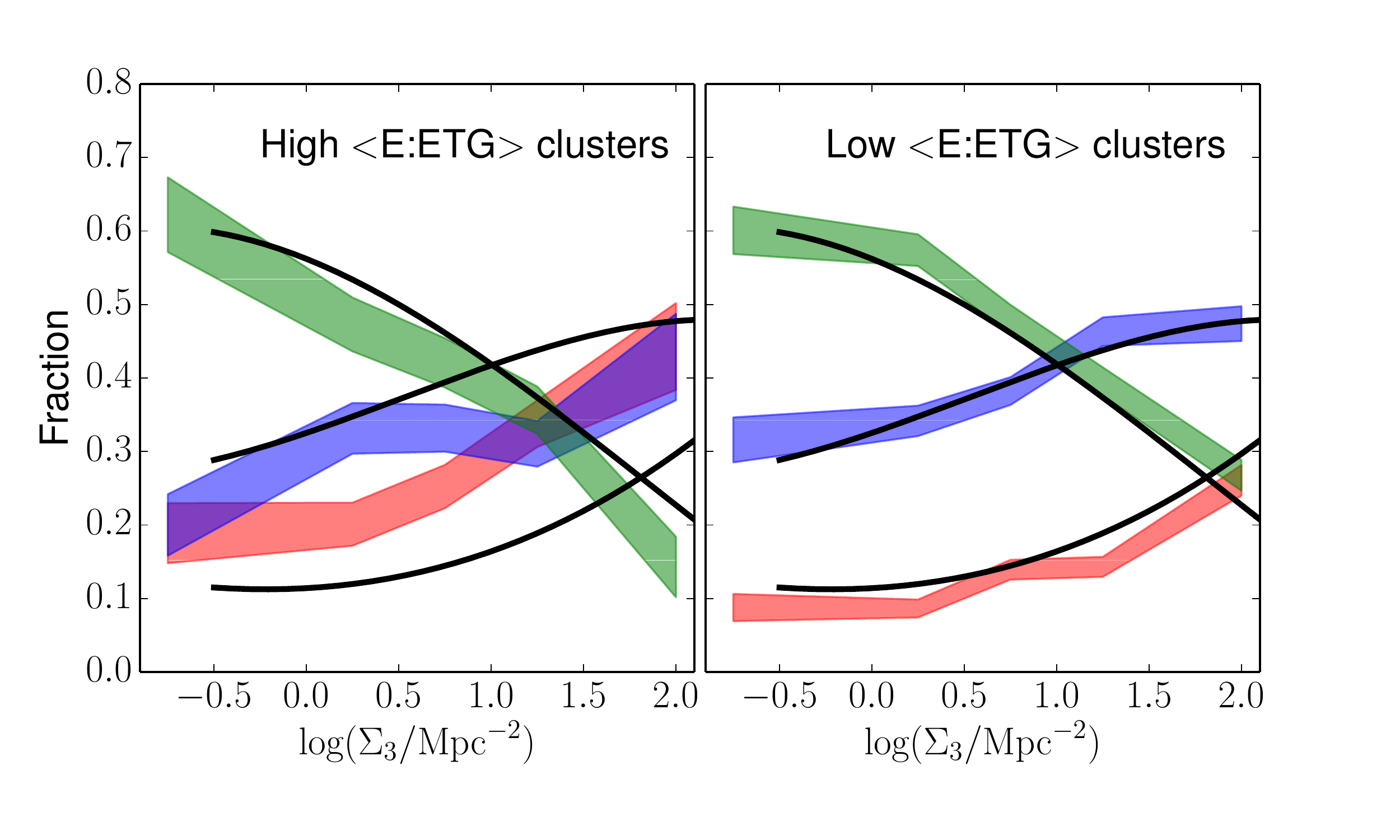} 
   \caption{\textbf{Left:} the \mdr relation for low-density clusters with high \EETG ratios. \textbf{Right:} the \mdr relation for low-density clusters with low \EETG ratios. \textbf{All:} Clusters with higher \EETG ratios appear to have similar \fESig and \fSSig relations while these relations diverge for clusters with low \EETG ratios. Note that the S+I fraction is similar in both cases. The black lines represent the polynomial fits in Fig. \ref{fig:D80_MDR}}
   \label{fig:HiLoEETGMDR}
\end{figure}

We show the \mdr relations for the high and low \EETG ratio clusters in Fig. \ref{fig:HiLoEETGMDR}. While the S+I fraction is similar in both cases, the \fESig and \fSSig relations appear to converge for clusters with higher \EETG ratios: the S0 fractions are lower and the E fractions higher.

\subsection{The \sigsig relation}

\begin{figure} 
   \centering
   \includegraphics[width=0.5\textwidth]{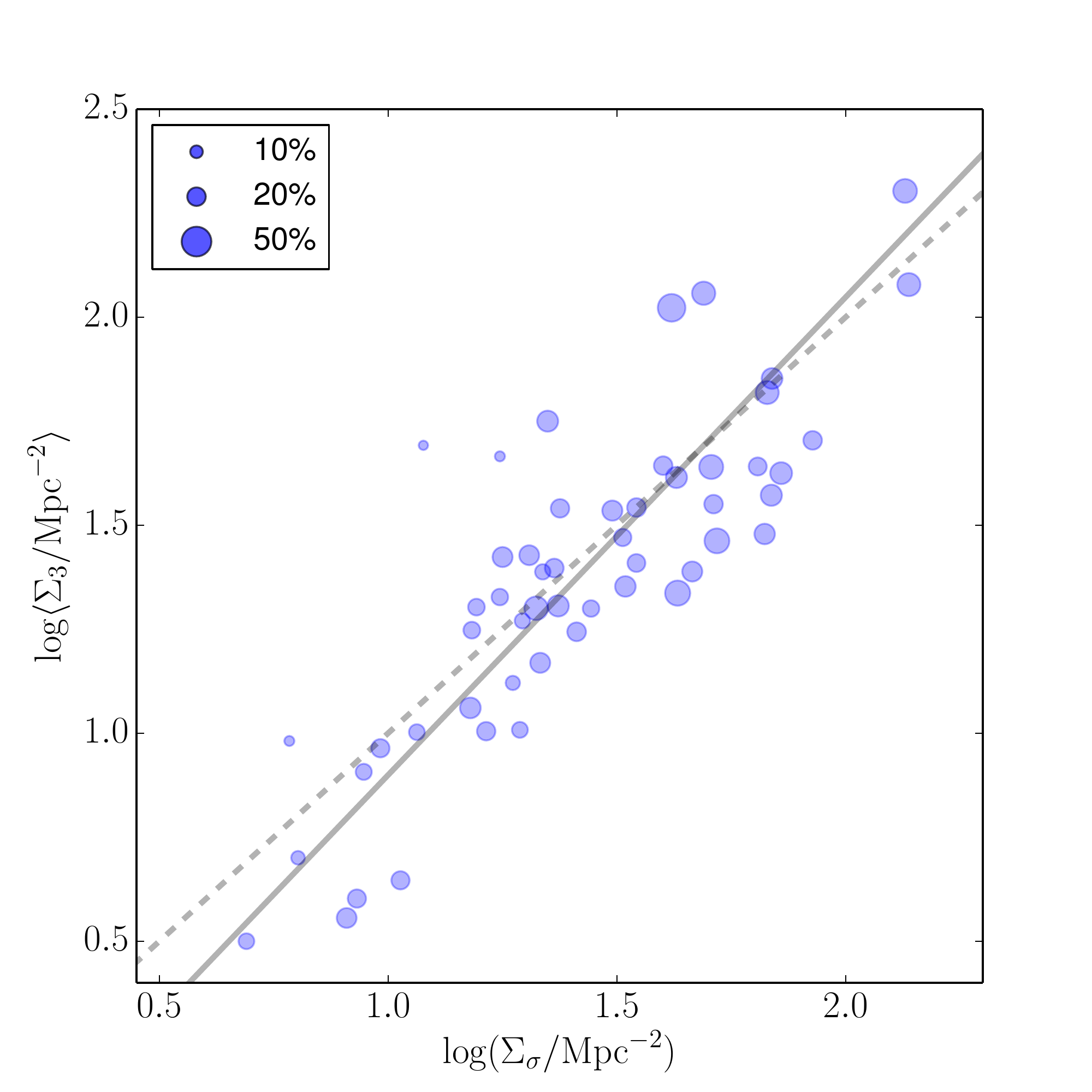} 
   \caption{The \sigsig relation, linking the average \emph{local} density of galaxies with the \emph{global} density in an adaptive circular aperture. Symbol sizes represent the number of cluster galaxies contained in the adaptive circular aperture. The dashed line is the 1:1 relation, while the solid line is a bisector fit to the data.}
   \label{fig:mLogSig-logAvDens}
\end{figure}

In Fig. \ref{fig:mLogSig-logAvDens}, we show the $\log\langle\Sigma_{3}\rangle$--$\log\Sigma_{\sigma}$ relation, demonstrating that the mean local density correlates well with the \emph{global} density of the central cluster regions. The size of the points represents the fraction of all cluster galaxies contained within the adaptive aperture: there are three clusters for which the number of galaxies contained in the adaptive aperture is exceptionally low ($<$10\%; Abell~548, Abell~2256, \& DC0326-53) as the median coordinates of the cluster galaxies do not adequately represent the centres of the clusters (or substructure in the clusters). Unsurprisingly they are Bautz-Morgan types III, II/III and III, respectively. If all galaxies were distributed Normally in a circularly symmetric distribution, we would expect 39\% of galaxies to be contained within our adaptive aperture. In practice, most (all but one) clusters contain fewer galaxies than this in the adaptive aperture, because cluster galaxies are neither Normally, nor circularly symmetrically distributed on the sky. 

We fit the data using a Bisector estimator \citep[in the absence uncertainties on either axis, ][]{Isobe1990}. While the link between average local density and global density is very close to the 1:1 relation (dashed line), the low-density clusters scatter below it, suggesting they are less well represented by a circular aperture.  Similarly, there is a tendency for global densities to be underestimated when fewer galaxies are contained within the adaptive aperture (at fixed \apld, symbols are smaller towards lower \aagd). Furthermore, clusters with higher global densities tend to have more galaxies in the adaptive aperture (at higher global densities, a larger fraction of the cluster galaxies are contained within the adaptive aperture).

\section{Discussion} \label{sec:discussion}

In the following section, we argue for the following: 
the \fESig relation is not driven just by $\log\Sigma_{3}$, but is also dependent on $\log\langle\Sigma_{3}\rangle$, which suggests a different mechanism is responsible for creating this relation;
in any given cluster, the \EETG ratio is constant at around 30\% which is consistent with a \SRETG fraction of 15\%;
the higher scatter in the \EETG ratio at low average densities is real, and clusters with high \EETG ratios all have high E-fractions;
the \sigsig relation gives meaning to our measure of average local density, and highlights our inability to directly measure an average global density. 
Finally we discuss the implications of these findings in the context of formation mechanisms for ETGs. 

\subsection{The \fESig relation is different to the \fSSig \& \fSISig relations}

The number fractions of S+I and S0 galaxies in Fig. \ref{fig:LMHMDR} appear to be independent of the $\log\langle\Sigma_{3}\rangle$: we see the same \fSSig and \fSISig relations for low-, medium- and high-density clusters. This is entirely consistent with the results of D80. 

However, the situation is more complicated for the Es. The original \mdr relation of D80 can be interpreted as a universal link between all morphologies and local projected density. Fig. \ref{fig:LMHMDR} shows that the relation between E-fraction and \lS3 is not that simple: the E-fraction depends on the \emph{relative} local density within the cluster. In low-density clusters, the E-fraction attains a similar peak value to the high-density clusters, but at appreciably lower local density. This peak in the E-fraction always occurs at the highest local densities available in a given cluster; that is, the highest \emph{relative} local densities. 

The results from the \kmdr relation are strikingly similar to the results presented here. In \citet{Houghton2013}, the SR:ETG ratio reached peak values at very different local densities for Virgo, Coma and Abell~1689. The similar behaviour of the SR:ETG and E:ETG ratios suggests that they are tracing the same, or at least a highly correlated, population. This is despite 66\% of the Es in the ATLAS3D survey being Fast Rotators, which are kinematically indistinguishable from S0s, and grouped with the majority of S0s in any \kmdr relations \citep{ATLAS3DIII,Houghton2013}. 

\subsection{The \EETG ratio is around 30\% in any given cluster}

Fig. \ref{fig:AvEETGfrac} shows that, while the ratio of star-forming to quiescent disks \SDISK strongly depends on $\log\langle\Sigma_{3}\rangle$, the \EETG ratio is indifferent to it: the \EETG ratio is always around 30\%. 

This behaviour is remarkably similar to that of the \SRETG ratio \citep{Houghton2013}. While the \SRETG ratio was found to be to be 15\%, here the \EETG ratio is 30\%. The \A3D project kinematically classified 253 ETGs and we can use these statistics to link kinematic morphologies with visual morphologies. Let us define the `probability of error' in morphological classifications, by assuming there is no error in kinematic classifications. 
Only 80\% (179/224) of the FRs were classified S0, so the probability of mistakenly classifying a ``true'' lenticular (i.e. an FR) as an E is $p(E | FR)=0.2$. Of all the SRs, 65\% (23/36) were classified as Es, so the probability of mistakenly classifying a ``true'' elliptical (i.e. an SR) as an S0 is $p(S0 | SR)=0.35$. Assuming ETGs are either Es or S0s (no misclassification of SRs or FRs as LTGs and \emph{vice versa}), we can infer $p(S0 | FR)$ and $p(E | SR)$, as $p(E | FR) + p(S0 | FR) = p(E | SR) + p(S0 | SR) = 1$. Using these probabilities we can then estimate the number of morphological Es and S0s given the true number of FRs and SRs, 
\begin{equation}
\label{eq:S0EfromFRSR}
\m{n(\textrm{S0})}{n(\textrm{E})} = {\bf P} \m{n(\textrm{FR})}{n(\textrm{SR})}
\end{equation}
where 
\begin{equation}
{\bf P} = \mm{p(\textrm{S0}|\textrm{FR})}{p(\textrm{S0}|\textrm{SR})}{p(\textrm{E}|\textrm{FR})}{p(\textrm{E}|\textrm{SR})}.
\end{equation}
We can also invert Eq. \ref{eq:S0EfromFRSR} to estimate the number of FRs and SRs from the number of morphological Es and S0s,
\begin{equation}
\label{eq:FRSRfromS0E}
\m{n(\textrm{FR})}{n(\textrm{SR})}  = {\bf P}^{-1} \m{n(\textrm{S0})}{n(\textrm{E})}
\end{equation}
Using these equations, we confirm that the apparent \EETG ratio would be 27\% if the \SRETG fraction is only 15\% (and \emph{visa versa}). Accordingly, we can appreciate why recent studies of the \kmdr have found that the \SRETG fraction remains roughly constant at 15\%; indeed, it could have been anticipated if the independence of the \EETG ratio had been known \emph{a priori}. 

\subsection{The highest E-fractions are found in lower-density clusters}

We can use Eq. \ref{eq:FRSRfromS0E} to estimate the true number of FRs and SRs in individual clusters. Of great interest are the clusters with high \EETG ratios, which are only found at low $\log\langle\Sigma_{3}\rangle$. In Fig. \ref{fig:AvEETGfrac}, we see that there are \nhighEfrac clusters with \EETG$>$40\% and $\log\langle\Sigma_{3}\rangle$$<$1.5 (there are \nlowEfrac clusters with \EETG$<$0.4 and  $\log\langle\Sigma_{3}\rangle$$<$1.5). Such high \EETG ratios are a result of much higher E-fractions and marginally lower S0-fractions: mean E-fractions are 25\% and 12\% for high and low \EETG clusters, respectively; mean S0-fractions are 32\% and 45\% for the high and low \EETG clusters, respectively. The mean S+I fractions are very similar for both: 40\% and 42\% for the high and low \EETG clusters, respectively. Consequently, it appears that the high \EETG ratios are predominantly driven by a doubling of the E-fractions. Note that \EETG ratio of 40\% suggest \SRETG ratio of 45\% using Eq. \ref{eq:FRSRfromS0E}; similarly an \EETG ratio of 50\% suggests an \SRETG ratio of almost 70\%. 

A key question to address is whether we have selected the upper envelope of a distribution (with large intrinsic scatter or large measurement uncertainty), or if these high \EETG clusters represent a genuinely different type of cluster, perhaps at a different stage of cluster evolution (e.g. unrelaxed or merging). The scale of the random uncertainties above and below $\log\langle\Sigma_{3}\rangle=1.5$ are comparable, so the high scatter at lower $\log\langle\Sigma_{3}\rangle$ does not originate from increased measurement uncertainty. In Fig. \ref{fig:AvEETGfrac}, there is an apparent gap between low \EETG clusters and high \EETG clusters below $\log\langle\Sigma_{3}\rangle$=1.5. With the present data, we are unable to determine if this represents a genuine split and bimodality in the \EETG ratio, or whether the distribution is continuous but sparsely sampled. 

In Fig. \ref{fig:HiLoEETGMDR}, we showed that the high \EETG clusters show a normal \fSISig relation, while the \fESig and \fSSig relations are offset from the usual trends and almost converge. This is not trivial: several studies of the \kmdr relation have shown that in unrelaxed clusters, the SRs are not always found in the densest regions \citep{Scott2014,Fogarty2014}. Furthermore, if the most massive galaxies (predominantly Es and/or SRs) merge in cluster mergers, then one might expect the high \EETG clusters to be unrelaxed and show irregular \mdr relations; this is not the case, they do show \fESig and \fSSig relations with higher numbers of ETGs in the denser environments. We therefore conclude that there is an \emph{intrinsic} variation in the morphological content of lower-density clusters: some have double the normal E fractions and presumably much higher SR fractions. 

\subsection{The significance of the \sigsig relation}

We have demonstrated that the E-fraction within clusters depends on the \emph{average} projected local density of galaxies in those clusters, while the average E-fraction in the ETG population is independent of it. However, it is not immediately obvious what our `average projected local density' measures. To address this problem, we investigated links between $\log\langle\Sigma_{3}\rangle$ and a \emph{global} measure of galaxy density. Fig. \ref{fig:mLogSig-logAvDens} illustrates that clusters with a higher average projected \emph{local} density also have a higher projected \emph{global} density. This is not immediately obvious, but equally perhaps not surprising. What is peculiar is that when we define low-, medium- and high-density clusters based on this global measure of density, we do \emph{not} recover the different \fESig relations seen in the lower panel of Fig. \ref{fig:LMHMDR}. Clearly, there is too much uncertainty in a global measure of density to recover these subtle changes.

Our method of determining a global density relies on defining a single circular aperture to encompass the cluster galaxies, but many clusters are not circularly symmetric on the sky. This is an important limitation of the global approach to measuring density. In Fig. \ref{fig:mLogSig-logAvDens}, at a given average \emph{local} density, there is a tendency for the global density to be lower when the circular aperture contained a smaller fraction of galaxies (i.e. the cluster was not well represented by the single centroid and circular aperture). This appears to be the dominant source of uncertainty (intrinsic scatter) in the \sigsig relation, and may explain why we don't recover the different \fESig relations when using global density.

Clusters frequently show multiple sub-structures and are not all circularly symmetric; for such `irregular' morphologies, the original use of local projected density in D80, and the use of average local density here, have lead to a better understanding of how morphology links to environment. While the \sigsig relation shows that there is a link between local environment and global environment, the significant scatter warns that we are not yet able to measure global environment reliably for all cluster morphologies.

The \sigsig relation shows some interesting features in its own right. There is a tendency for denser clusters to have a larger fraction of galaxies in the adaptive aperture, suggesting they are either more circularly symmetric and/or more concentrated in their galaxy distributions. The fact that there is a \sigsig relation -- a link between local density and global density -- is not immediately obvious. For comparison, recall that more massive galaxies do not necessarily have higher stellar densities \citep[see Fig. 1 of ][]{ATLAS3DXX}. It is not clear what mechanism could be responsible for this relation: in a hierarchical framework, when forming progressively more massive clusters from more massive sub-units, there may be a process which dissipates the orbital angular momentum to increase the density of the galaxies, such as dynamical friction. Equally, as the most massive over-densities are the first to collapse, higher local densities in massive clusters may reflect the higher densities of galaxies at the formation epoch; lower mass clusters may only have virialised recently, after considerable expansion of the Universe. Both these possibilities rely on a link between global density and mass.

\subsection{Implications for formation mechanisms of Es and S0s}

The transformation of spirals into lenticulars is a common explanation for the fall/rise in the S+I/S0 fractions with local density. However, the details of such a mechanism are far from certain and there are a number of contenders \citep{BoselliGavazzi2006}. Recent evidence suggests that the majority of cluster S0s were ``pre-processed'' and quenched in the group stages, before entering a cluster environment \citep[e.g.][]{Mihos2004,Moran2007}, via galaxy-galaxy interactions such as tidal interactions, harassment \citep{Moore1996} or gravitational heating \citep{KhochfarOstriker2008}. Indeed, the \kmdr relation of \citet{Cappellari2012} probes mainly field and group environments: as they highlight, finding such a tight correlation in these lower-density environments suggests that the \mdr relation is initiated in the group stages. Similarly, \citet{HelsdonPonman2003} find that the morphological segregation in X-ray bright groups is similar to that of clusters, further supporting the suggestion that the group environment is mostly responsible for the \mdr relations. \citet{Dressler2013} provide further support for pre-processing: they studied the stellar populations and substructures in intermediate redshift clusters and found a correlation between the passive and post-starburst fraction and the mass of the substructures, indicating the substructure environment (and not the overall cluster environment) dictates the quenching efficiency. Galaxy-galaxy interactions in the group stage are also far more likely to be related to \emph{local} number density, which may explain its power in constructing the \mdr relations. Cluster-scale effects like ram-pressure stripping \citep{GunnGott1972}, although evident \citep{Chung2009}, are not commonly thought to be the dominant route for creating the cluster S0 population and the \mdr relation (indeed, this was also the original conclusion of D80).

So where do Es fit in to this picture? What distinguishes Es from S0s morphologically, is the complete absence of a disk. One can imagine numerous scenarios where a disk is destroyed. Historically, major mergers \citep{Holmberg1941,ToomreToomre1972} were commonly used to explain the assembly of Es \citep{Zwicky1959,Barnes1988,BarnesHernquist1992}. But the major-merger scenario is not without its problems: unless collisions are head-on or the internal angular momentum of the stars aligned to cancel the orbital angular momentum of the merger, the remnants rotate too quickly and are too flattened to mimic slowly rotating elliptical galaxies \citep{White1979a,White1979b,ATLAS3DVI}\footnote{Note that the simulations of \citet{ATLAS3DVI} included a dark matter halo, while those of \citet{White1979a,White1979b} did not.}. A (multiple) minor merger scenario is now more commonly invoked to generate the low angular momentum state of the slowly rotating ellipticals \citep{ATLAS3DIIX}. While this scenario has an appealing synergy in explaining the apparent size evolution of ETGs \citep[e.g.][]{Trujillo2007,Naab2009}, such a protracted assembly over time from many low-mass systems seems difficult to square with the high metallicity and alpha-enhancement of the stellar populations in massive galaxies \citep[implying high redshift star formation over short timescales, ][]{Thomas2005}, not to mention the abundance of Es at high redshift \citep{Dressler1997,Postman2005,Smith2005}. 

We have shown that Es obey a fundamentally different relation with local density compared to the disk systems (S0s and S+Is): they reside in regions of the highest \emph{relative} density. They either formed there, or accumulated there via some process. Take for example the Coma cluster: two super-giant, massive Es inhabit the centre of this cluster and they \emph{will} merge in several Gyrs \citep{Gerhard2007}. These galaxies didn't form in their present locations: dynamical friction from the cluster-scale dark matter halo has played a role in dissipating angular momentum to bring these two super-giant Es together. Significant sub-structure (i.e. galaxies) is associated with each of these super-giant Es and as in the general 3-body problem, the lower mass sub-structures may help transfer angular momentum away from the two most massive galaxies by being ejected. This will advance the merger, as may tidal effects. But the ejection of substructures during the merger process will result in less of the initial orbital angular becoming locked into the remnant, and a lower stellar angular momentum state (a slow rotator). In addition, the substructure ejected on radial orbits could return, being stripped by tidal effects and even resulting in minor mergers to further lower the net angular momentum. But for the vast majority of cluster galaxies, mergers are rare \citep[note merger signatures are long-lived, ][]{Sheen2012,Yi2013}. However, harassment or high-velocity encounters \citep{Moore1996} are common. All that is required to transform an S0 to a E is destruction of the visible disk. Tidal effects and harassment may have the power to do this. The fall in the S0-fraction at the highest densities in Fig. \ref{fig:LMHMDR} coincides with the peaks in the E-fractions, suggestion S0s are being transformed into Es there. Furthermore, \citet{DEugenio2015} have found that the ellipticities of passive galaxies decrease towards the centres of clusters, suggesting that the disks may be being disrupted or thickened as they fall into the denser regions of the clusters. Thus while the \emph{central} SRs may form in the group stages, the remaining Es may have `formed' (assumed their current morphology, previously being S0s) in the cluster itself.

In \citet{Dressler1997}, studies of intermediate redshift ($z\sim0.5$) clusters yielded similar or higher E-fractions, with a deficit of S0s and an abundance of Ss. Indeed, typical \EETG ratios at $z\sim0.5$ were as high as 75\%, far higher than even the most E-rich clusters at low-z. Similar results were found at z$\sim$1, with an increased ETG-fraction in denser environments but similar ETG-fractions in the field \citep{Smith2005,Postman2005}. The conclusion from these observations is that the formation epoch of cluster Es is much earlier than cluster S0s, which formed later from quenched spirals. Accordingly, the clusters we see at low-z showing the highest \EETG ratios in Fig. \ref{fig:AvEETGfrac} may be relics of the earlier Universe which have undergone slower evolution to the present day. However, the implicit assumption that Es cannot be transformed into S0s (and \emph{visa versa}) conflicts with the formation histories in \citet{ATLAS3DXXV}, where many SRs (admittedly only 64\% Es) are `formed' (i.e. transform into their current kinematic morphology) at z$<1$. Given that transformations between \emph{kinematic} morphologies are seen in both directions in the simulations (FR $\leftrightarrow$ SR), transformations between visual morphologies may also be possible. 

Kinematic \emph{and} visual morphology-density relations for intermediate redshift clusters may help us better understand this problem. The kinematic imprint of a disk is longer lived than the visible disk itself: comparing the S0 and FR fractions at intermediate and low redshift will reveal if S0s are being transformed into (fast rotating) Es: a fall in the S0 fraction and a rise in the E-fraction at higher densities \emph{for just the FR population} would be interesting indeed. There is a school-of-thought that the high-redshift Universe was completely different and substantially more chaotic to the one we see locally. With mergers orders of magnitude more frequent \citep{Fakhouri2010}, star formation greatly enhanced \citep{Madau1996} and AGN activity similarly boosted \citep{Cowie2003}, we might well expect to create more dispersion-dominated structures with no stable disk. The same \emph{combined} study of the \kmdr and \mdr relations could determine whether Es (or SRs) were really formed \emph{in their present structures} at higher redshifts; or if the central galaxies in clusters gradually form their low angular momentum state from minor mergers. We can further determine if the statistics of our visual morphologies at higher redshift tally with those at low redshift, using the S0:FR and E:SR ratios, which is crucial given the doubts raised by \citet{Andreon1998} and \citet{Holden2009}.

\section{Conclusions}
\label{sec:conclusion}
We revisited the original data of \citet{Dressler1980cat}, updating the cluster redshifts and projected densities. We use the \emph{average} local projected density to characterise clusters into low-, medium- and high-density classes and show that:
\begin{itemize}
\item the rise in elliptical fraction with local density for low-density clusters occurs at lower densities than the same rise for high-density clusters; thus, the distribution of elliptical galaxies within a cluster depends on the \emph{relative} local projected density of galaxies, which is not the case for the lenticular or late-type fractions; 
\item the \EETG ratio in a cluster is independent of the average local density of galaxies in that cluster, with a constant value of around 30\%; in the fast and slow rotator paradigm, this is consistent with a constant \SRETG fraction of 15\%; conversely, the \SDISK ratio varies strongly with average local density;
\item contrary to the canonical \mdr relation, clusters with the highest \EETG ratios (and overall elliptical fractions) are found in lower-density clusters; the \mdr relations for these high \EETG ratio clusters show a convergence of the \fESig and \fSSig relations;
\item there is a simple relation between the average local projected density of galaxies and the global density of galaxies in a cluster, such that clusters with a higher average local density have a higher global density; however our global measure of density remains crude and poorly characterises irregular cluster morphologies.  
\end{itemize}

\section*{Acknowledgments}
We thank the referee, Alan Dressler, for his comments and suggestions which considerably improved this work. We gratefully acknowledge interesting discussions with Roger Davies and Francesco D'Eugenio and thank them for reading over earlier drafts of this manuscript. RCWH was supported by the Science and Technology Facilities Council
[STFC grant numbers ST/H002456/1 \& ST/K00106X/1].
This research has made use of the NASA/IPAC Extragalactic Database (NED) which is operated by the Jet Propulsion Laboratory, California Institute of Technology, under contract with the National Aeronautics and Space Administration.
This research has made use of the VizieR catalogue access tool, CDS, Strasbourg, France. The original description of the VizieR service was published in A\&AS 143, 23.

\bibliographystyle{mn2e}
\bibliography{MASTERBIB}

\begin{thebibliography}{}

\bibitem[\protect\citeauthoryear{{Andreon}}{{Andreon}}{1998}]{Andreon1998}
{Andreon} S.,  1998, \apj, 501, 533

\bibitem[\protect\citeauthoryear{{Bacon}, {Copin}, {Monnet}, {Miller},
  {Allington-Smith}, {Bureau}, {Carollo}, {Davies} \& {et al.}}{{Bacon}
  et~al.}{2001}]{SAURONI}
{Bacon} R.,  {Copin} Y.,  {Monnet} G.,  {Miller} B.~W.,  {Allington-Smith}
  J.~R.,  {Bureau} M.,  {Carollo} C.~M.,  {Davies}   {et al.} 2001, \mnras,
  326, 23

\bibitem[\protect\citeauthoryear{{Barnes}}{{Barnes}}{1988}]{Barnes1988}
{Barnes} J.~E.,  1988, \apj, 331, 699

\bibitem[\protect\citeauthoryear{{Barnes} \& {Hernquist}}{{Barnes} \&
  {Hernquist}}{1992}]{BarnesHernquist1992}
{Barnes} J.~E.,  {Hernquist} L.,  1992, \araa, 30, 705

\bibitem[\protect\citeauthoryear{{Bois}, {Emsellem}, {Bournaud}, {Alatalo},
  {Blitz}, {Bureau}, {Cappellari}, {Davies} \& {et al.}}{{Bois}
  et~al.}{2011}]{ATLAS3DVI}
{Bois} M.,  {Emsellem} E.,  {Bournaud} F.,  {Alatalo} K.,  {Blitz} L.,
  {Bureau} M.,  {Cappellari} M.,  {Davies} R.~L.,    {et al.} 2011, \mnras,
  416, 1654

\bibitem[\protect\citeauthoryear{{Boselli} \& {Gavazzi}}{{Boselli} \&
  {Gavazzi}}{2006}]{BoselliGavazzi2006}
{Boselli} A.,  {Gavazzi} G.,  2006, \pasp, 118, 517

\bibitem[\protect\citeauthoryear{{Cappellari}, {Emsellem}, {Bacon}, {Bureau},
  {Davies}, {de Zeeuw}, {Falc{\'o}n-Barroso}, {Krajnovi{\'c}}, {Kuntschner},
  {McDermid}, {Peletier}, {Sarzi}, {van den Bosch} \& {van de
  Ven}}{{Cappellari} et~al.}{2007}]{SAURONX}
{Cappellari} M.,  {Emsellem} E.,  {Bacon} R.,  {Bureau} M.,  {Davies} R.~L.,
  {de Zeeuw} P.~T.,  {Falc{\'o}n-Barroso} J.,  {Krajnovi{\'c}} D.,
  {Kuntschner} H.,  {McDermid} R.~M.,  {Peletier} R.~F.,  {Sarzi} M.,  {van den
  Bosch} R.~C.~E.,    {van de Ven} G.,  2007, \mnras, 379, 418

\bibitem[\protect\citeauthoryear{{Cappellari}, {Emsellem}, {Krajnovi{\'c}},
  {McDermid}, {Scott}, {Verdoes Kleijn}, {Young}, {Alatalo} \& {et
  al.}}{{Cappellari} et~al.}{2011}]{ATLAS3DI}
{Cappellari} M.,  {Emsellem} E.,  {Krajnovi{\'c}} D.,  {McDermid} R.~M.,
  {Scott} N.,  {Verdoes Kleijn} G.~A.,  {Young} L.~M.,  {Alatalo} K.,    {et
  al.} 2011, \mnras, 413, 813

\bibitem[\protect\citeauthoryear{{Cappellari}, {Emsellem}, {Krajnovi{\'c}},
  {McDermid}, {Serra}, {Alatalo} \& {et al.}}{{Cappellari}
  et~al.}{2011}]{ATLAS3DVII}
{Cappellari} M.,  {Emsellem} E.,  {Krajnovi{\'c}} D.,  {McDermid} R.~M.,
  {Serra} P.,  {Alatalo} K.,    {et al.} 2011, \mnras, 416, 1680

\bibitem[\protect\citeauthoryear{{Cappellari}, {McDermid}, {Alatalo}, {Blitz},
  {Bois}, {Bournaud}, {Bureau}, {Crocker} \& {et al.}}{{Cappellari}
  et~al.}{2013}]{ATLAS3DXX}
{Cappellari} M.,  {McDermid} R.~M.,  {Alatalo} K.,  {Blitz} L.,  {Bois} M.,
  {Bournaud} F.,  {Bureau} M.,  {Crocker} A.~F.,    {et al.} 2013, \mnras, 432,
  1862

\bibitem[\protect\citeauthoryear{{Cappellari}, {McDermid}, {Alatalo}, {Blitz},
  {Bois}, {Bournaud} \& {et al.}}{{Cappellari} et~al.}{2012}]{Cappellari2012}
{Cappellari} M.,  {McDermid} R.~M.,  {Alatalo} K.,  {Blitz} L.,  {Bois} M.,
  {Bournaud} F.,    {et al.} 2012, \nat, 484, 485

\bibitem[\protect\citeauthoryear{{Chung}, {van Gorkom}, {Kenney}, {Crowl} \&
  {Vollmer}}{{Chung} et~al.}{2009}]{Chung2009}
{Chung} A.,  {van Gorkom} J.~H.,  {Kenney} J.~D.~P.,  {Crowl} H.,    {Vollmer}
  B.,  2009, \aj, 138, 1741

\bibitem[\protect\citeauthoryear{{Cowie}, {Barger}, {Bautz}, {Brandt} \&
  {Garmire}}{{Cowie} et~al.}{2003}]{Cowie2003}
{Cowie} L.~L.,  {Barger} A.~J.,  {Bautz} M.~W.,  {Brandt} W.~N.,    {Garmire}
  G.~P.,  2003, \apjl, 584, L57

\bibitem[\protect\citeauthoryear{{Davis} \& {Geller}}{{Davis} \&
  {Geller}}{1976}]{DavisGeller76}
{Davis} M.,  {Geller} M.~J.,  1976, \apj, 208, 13

\bibitem[\protect\citeauthoryear{{D'Eugenio}, {Houghton}, {Davies} \&
  {Bont{\`a}}}{{D'Eugenio} et~al.}{2013}]{DEugenio2013}
{D'Eugenio} F.,  {Houghton} R.~C.~W.,  {Davies} R.~L.,    {Bont{\`a}} E.~D.,
  2013, \mnras, 429, 1258

\bibitem[\protect\citeauthoryear{{D'Eugenio}, {Houghton}, {Davies} \& {Dalla
  Bont{\`a}}}{{D'Eugenio} et~al.}{2015}]{DEugenio2015}
{D'Eugenio} F.,  {Houghton} R.~C.~W.,  {Davies} R.~L.,    {Dalla Bont{\`a}} E.,
   2015, ArXiv e-prints

\bibitem[\protect\citeauthoryear{{Dressler}}{{Dressler}}{1980a}]{Dressler1980cat}
{Dressler} A.,  1980a, \apjs, 42, 565

\bibitem[\protect\citeauthoryear{{Dressler}}{{Dressler}}{1980b}]{Dressler1980}
{Dressler} A.,  1980b, \apj, 236, 351

\bibitem[\protect\citeauthoryear{{Dressler}, {Oemler} Jr., {Couch}, {Smail},
  {Ellis}, {Barger}, {Butcher}, {Poggianti} \& {Sharples}}{{Dressler}
  et~al.}{1997}]{Dressler1997}
{Dressler} A.,  {Oemler} Jr. A.,  {Couch} W.~J.,  {Smail} I.,  {Ellis} R.~S.,
  {Barger} A.,  {Butcher} H.,  {Poggianti} B.~M.,    {Sharples} R.~M.,  1997,
  \apj, 490, 577

\bibitem[\protect\citeauthoryear{{Dressler}, {Oemler} Jr., {Poggianti},
  {Gladders}, {Abramson} \& {Vulcani}}{{Dressler} et~al.}{2013}]{Dressler2013}
{Dressler} A.,  {Oemler} Jr. A.,  {Poggianti} B.~M.,  {Gladders} M.~D.,
  {Abramson} L.,    {Vulcani} B.,  2013, \apj, 770, 62

\bibitem[\protect\citeauthoryear{{Dressler} \& {Shectman}}{{Dressler} \&
  {Shectman}}{1988}]{DresslerSchectman1988}
{Dressler} A.,  {Shectman} S.~A.,  1988, \aj, 95, 284

\bibitem[\protect\citeauthoryear{{Emsellem}, {Cappellari}, {Krajnovi{\'c}},
  {Alatalo}, {Blitz}, {Bois}, {Bournaud}, {Bureau} \& {et al.}}{{Emsellem}
  et~al.}{2011}]{ATLAS3DIII}
{Emsellem} E.,  {Cappellari} M.,  {Krajnovi{\'c}} D.,  {Alatalo} K.,  {Blitz}
  L.,  {Bois} M.,  {Bournaud} F.,  {Bureau} M.,    {et al.} 2011, \mnras, 414,
  888

\bibitem[\protect\citeauthoryear{{Fakhouri}, {Ma} \&
  {Boylan-Kolchin}}{{Fakhouri} et~al.}{2010}]{Fakhouri2010}
{Fakhouri} O.,  {Ma} C.-P.,    {Boylan-Kolchin} M.,  2010, \mnras, 406, 2267

\bibitem[\protect\citeauthoryear{{Fogarty}, {Scott}, {Owers}, {Brough},
  {Croom}, {Pracy}, {Houghton}, {Bland-Hawthorn} \& {et al.}}{{Fogarty}
  et~al.}{2014}]{Fogarty2014}
{Fogarty} L.~M.~R.,  {Scott} N.,  {Owers} M.~S.,  {Brough} S.,  {Croom} S.~M.,
  {Pracy} M.~B.,  {Houghton} R.~C.~W.,  {Bland-Hawthorn} J.,    {et al.} 2014,
  \mnras, 443, 485

\bibitem[\protect\citeauthoryear{{Gerhard}, {Arnaboldi}, {Freeman}, {Okamura},
  {Kashikawa} \& {Yasuda}}{{Gerhard} et~al.}{2007}]{Gerhard2007}
{Gerhard} O.,  {Arnaboldi} M.,  {Freeman} K.~C.,  {Okamura} S.,  {Kashikawa}
  N.,    {Yasuda} N.,  2007, \aap, 468, 815

\bibitem[\protect\citeauthoryear{{Gunn} \& {Gott} III}{{Gunn} \&
  {Gott}}{1972}]{GunnGott1972}
{Gunn} J.~E.,  {Gott} III J.~R.,  1972, \apj, 176, 1

\bibitem[\protect\citeauthoryear{{Helsdon} \& {Ponman}}{{Helsdon} \&
  {Ponman}}{2003}]{HelsdonPonman2003}
{Helsdon} S.~F.,  {Ponman} T.~J.,  2003, \mnras, 339, L29

\bibitem[\protect\citeauthoryear{{Holden}, {Franx}, {Illingworth}, {Postman},
  {van der Wel}, {Kelson}, {Blakeslee}, {Ford}, {Demarco} \& {Mei}}{{Holden}
  et~al.}{2009}]{Holden2009}
{Holden} B.~P.,  {Franx} M.,  {Illingworth} G.~D.,  {Postman} M.,  {van der
  Wel} A.,  {Kelson} D.~D.,  {Blakeslee} J.~P.,  {Ford} H.,  {Demarco} R.,
  {Mei} S.,  2009, \apj, 693, 617

\bibitem[\protect\citeauthoryear{{Holmberg}}{{Holmberg}}{1941}]{Holmberg1941}
{Holmberg} E.,  1941, \apj, 94, 385

\bibitem[\protect\citeauthoryear{{Houghton}, {Davies}, {D'Eugenio}, {Scott},
  {Thatte}, {Clarke}, {Tecza}, {Salter}, {Fogarty} \& {Goodsall}}{{Houghton}
  et~al.}{2013}]{Houghton2013}
{Houghton} R.~C.~W.,  {Davies} R.~L.,  {D'Eugenio} F.,  {Scott} N.,  {Thatte}
  N.,  {Clarke} F.,  {Tecza} M.,  {Salter} G.~S.,  {Fogarty} L.~M.~R.,
  {Goodsall} T.,  2013, \mnras, 436, 19

\bibitem[\protect\citeauthoryear{{Huchra}, {Macri}, {Masters}, {Jarrett},
  {Berlind}, {Calkins}, {Crook}, {Cutri} \& {et al.}}{{Huchra}
  et~al.}{2012}]{TMZ}
{Huchra} J.~P.,  {Macri} L.~M.,  {Masters} K.~L.,  {Jarrett} T.~H.,  {Berlind}
  P.,  {Calkins} M.,  {Crook} A.~C.,  {Cutri} R.,    {et al.} 2012, \apjs, 199,
  26

\bibitem[\protect\citeauthoryear{{Isobe}, {Feigelson}, {Akritas} \&
  {Babu}}{{Isobe} et~al.}{1990}]{Isobe1990}
{Isobe} T.,  {Feigelson} E.~D.,  {Akritas} M.~G.,    {Babu} G.~J.,  1990, \apj,
  364, 104

\bibitem[\protect\citeauthoryear{{Jones}, {Saunders}, {Colless}, {Read},
  {Parker}, {Watson}, {Campbell}, {Burkey} \& {et al.}}{{Jones}
  et~al.}{2004}]{6dF}
{Jones} D.~H.,  {Saunders} W.,  {Colless} M.,  {Read} M.~A.,  {Parker} Q.~A.,
  {Watson} F.~G.,  {Campbell} L.~A.,  {Burkey} D.,    {et al.} 2004, \mnras,
  355, 747

\bibitem[\protect\citeauthoryear{{Khochfar}, {Emsellem}, {Serra}, {Bois},
  {Alatalo}, {Bacon}, {Blitz}, {Bournaud} \& {et al.}}{{Khochfar}
  et~al.}{2011a}]{ATLAS3DVIII}
{Khochfar} S.,  {Emsellem} E.,  {Serra} P.,  {Bois} M.,  {Alatalo} K.,  {Bacon}
  R.,  {Blitz} L.,  {Bournaud} F.,    {et al.} 2011a, \mnras, 417, 845

\bibitem[\protect\citeauthoryear{{Khochfar}, {Emsellem}, {Serra}, {Bois},
  {Alatalo}, {Bacon}, {Blitz}, {Bournaud} \& {et al.}}{{Khochfar}
  et~al.}{2011b}]{ATLAS3DIIX}
{Khochfar} S.,  {Emsellem} E.,  {Serra} P.,  {Bois} M.,  {Alatalo} K.,  {Bacon}
  R.,  {Blitz} L.,  {Bournaud} F.,    {et al.} 2011b, \mnras, 417, 845

\bibitem[\protect\citeauthoryear{{Khochfar} \& {Ostriker}}{{Khochfar} \&
  {Ostriker}}{2008}]{KhochfarOstriker2008}
{Khochfar} S.,  {Ostriker} J.~P.,  2008, \apj, 680, 54

\bibitem[\protect\citeauthoryear{{Komatsu}, {Smith}, {Dunkley}, {Bennett},
  {Gold}, {Hinshaw}, {Jarosik}, {Larson} \& {et al.}}{{Komatsu}
  et~al.}{2011}]{WMAP7COSMO}
{Komatsu} E.,  {Smith} K.~M.,  {Dunkley} J.,  {Bennett} C.~L.,  {Gold} B.,
  {Hinshaw} G.,  {Jarosik} N.,  {Larson} D.,    {et al.} 2011, \apjs, 192, 18

\bibitem[\protect\citeauthoryear{{Madau}, {Ferguson}, {Dickinson},
  {Giavalisco}, {Steidel} \& {Fruchter}}{{Madau} et~al.}{1996}]{Madau1996}
{Madau} P.,  {Ferguson} H.~C.,  {Dickinson} M.~E.,  {Giavalisco} M.,  {Steidel}
  C.~C.,    {Fruchter} A.,  1996, \mnras, 283, 1388

\bibitem[\protect\citeauthoryear{{Mihos}}{{Mihos}}{2004}]{Mihos2004}
{Mihos} J.~C.,  2004, Clusters of Galaxies: Probes of Cosmological Structure
  and Galaxy Evolution, p.~277

\bibitem[\protect\citeauthoryear{{Moore}, {Katz}, {Lake}, {Dressler} \&
  {Oemler}}{{Moore} et~al.}{1996}]{Moore1996}
{Moore} B.,  {Katz} N.,  {Lake} G.,  {Dressler} A.,    {Oemler} A.,  1996,
  \nat, 379, 613

\bibitem[\protect\citeauthoryear{{Moran}, {Ellis}, {Treu}, {Smith}, {Rich} \&
  {Smail}}{{Moran} et~al.}{2007}]{Moran2007}
{Moran} S.~M.,  {Ellis} R.~S.,  {Treu} T.,  {Smith} G.~P.,  {Rich} R.~M.,
  {Smail} I.,  2007, \apj, 671, 1503

\bibitem[\protect\citeauthoryear{{Moretti}, {Poggianti}, {Fasano}, {Bettoni},
  {D'Onofrio}, {Fritz}, {Cava}, {Varela} \& {et al.}}{{Moretti}
  et~al.}{2014}]{WINGSdata}
{Moretti} A.,  {Poggianti} B.~M.,  {Fasano} G.,  {Bettoni} D.,  {D'Onofrio} M.,
   {Fritz} J.,  {Cava} A.,  {Varela} J.,    {et al.} 2014, \aap, 564, A138

\bibitem[\protect\citeauthoryear{{Naab}, {Johansson} \& {Ostriker}}{{Naab}
  et~al.}{2009}]{Naab2009}
{Naab} T.,  {Johansson} P.~H.,    {Ostriker} J.~P.,  2009, \apjl, 699, L178

\bibitem[\protect\citeauthoryear{{Naab}, {Oser}, {Emsellem}, {Cappellari},
  {Krajnovi{\'c}}, {McDermid}, {Alatalo}, {Bayet} \& {et al.}}{{Naab}
  et~al.}{2014}]{ATLAS3DXXV}
{Naab} T.,  {Oser} L.,  {Emsellem} E.,  {Cappellari} M.,  {Krajnovi{\'c}} D.,
  {McDermid} R.~M.,  {Alatalo} K.,  {Bayet} E.,    {et al.} 2014, \mnras, 444,
  3357

\bibitem[\protect\citeauthoryear{{Naab} \& {Trujillo}}{{Naab} \&
  {Trujillo}}{2006}]{NaabTrujillo2006}
{Naab} T.,  {Trujillo} I.,  2006, \mnras, 369, 625

\bibitem[\protect\citeauthoryear{{Oemler} Jr.}{{Oemler}}{1974}]{Oemler74}
{Oemler} Jr. A.,  1974, \apj, 194, 1

\bibitem[\protect\citeauthoryear{{Postman}, {Franx}, {Cross}, {Holden}, {Ford},
  {Illingworth}, {Goto}, {Demarco} \& {et al.}}{{Postman}
  et~al.}{2005}]{Postman2005}
{Postman} M.,  {Franx} M.,  {Cross} N.~J.~G.,  {Holden} B.,  {Ford} H.~C.,
  {Illingworth} G.~D.,  {Goto} T.,  {Demarco} R.,    {et al.} 2005, \apj, 623,
  721

\bibitem[\protect\citeauthoryear{{Rines}, {Geller}, {Kurtz} \&
  {Diaferio}}{{Rines} et~al.}{2003}]{Rines2003}
{Rines} K.,  {Geller} M.~J.,  {Kurtz} M.~J.,    {Diaferio} A.,  2003, \aj, 126,
  2152

\bibitem[\protect\citeauthoryear{{Scott}, {Davies}, {Houghton}, {Cappellari},
  {Graham} \& {Pimbblet}}{{Scott} et~al.}{2014}]{Scott2014}
{Scott} N.,  {Davies} R.~L.,  {Houghton} R.~C.,  {Cappellari} M.,  {Graham}
  A.~W.,    {Pimbblet} K.~A.,  2014, ArXiv e-prints

\bibitem[\protect\citeauthoryear{{Sheen}, {Yi}, {Ree} \& {Lee}}{{Sheen}
  et~al.}{2012}]{Sheen2012}
{Sheen} Y.-K.,  {Yi} S.~K.,  {Ree} C.~H.,    {Lee} J.,  2012, \apjs, 202, 8

\bibitem[\protect\citeauthoryear{{Smith}, {Treu}, {Ellis}, {Moran} \&
  {Dressler}}{{Smith} et~al.}{2005}]{Smith2005}
{Smith} G.~P.,  {Treu} T.,  {Ellis} R.~S.,  {Moran} S.~M.,    {Dressler} A.,
  2005, \apj, 620, 78

\bibitem[\protect\citeauthoryear{{Thomas}, {Maraston}, {Bender} \& {Mendes de
  Oliveira}}{{Thomas} et~al.}{2005}]{Thomas2005}
{Thomas} D.,  {Maraston} C.,  {Bender} R.,    {Mendes de Oliveira} C.,  2005,
  \apj, 621, 673

\bibitem[\protect\citeauthoryear{{Toomre} \& {Toomre}}{{Toomre} \&
  {Toomre}}{1972}]{ToomreToomre1972}
{Toomre} A.,  {Toomre} J.,  1972, \apj, 178, 623

\bibitem[\protect\citeauthoryear{{Trujillo}, {Conselice}, {Bundy}, {Cooper},
  {Eisenhardt} \& {Ellis}}{{Trujillo} et~al.}{2007}]{Trujillo2007}
{Trujillo} I.,  {Conselice} C.~J.,  {Bundy} K.,  {Cooper} M.~C.,  {Eisenhardt}
  P.,    {Ellis} R.~S.,  2007, \mnras, 382, 109

\bibitem[\protect\citeauthoryear{{White}}{{White}}{1979a}]{White1979a}
{White} S.~D.~M.,  1979a, \apjl, 229, L9

\bibitem[\protect\citeauthoryear{{White}}{{White}}{1979b}]{White1979b}
{White} S.~D.~M.,  1979b, \mnras, 189, 831

\bibitem[\protect\citeauthoryear{{Whitmore}, {Gilmore} \& {Jones}}{{Whitmore}
  et~al.}{1993}]{Whitmore1993}
{Whitmore} B.~C.,  {Gilmore} D.~M.,    {Jones} C.,  1993, \apj, 407, 489

\bibitem[\protect\citeauthoryear{{Yi}, {Lee}, {Jung}, {Ji} \& {Sheen}}{{Yi}
  et~al.}{2013}]{Yi2013}
{Yi} S.~K.,  {Lee} J.,  {Jung} I.,  {Ji} I.,    {Sheen} Y.-K.,  2013, \aap,
  554, A122

\bibitem[\protect\citeauthoryear{{Zwicky}}{{Zwicky}}{1959}]{Zwicky1959}
{Zwicky} F.,  1959, Handbuch der Physik, 53, 373

\end{thebibliography}

\appendix
\section{Projection effects in Abell~838, Abell~1631 \& DC2349-28}

While updating the cluster redshifts using the latest values from NED\footnote{http://ned.ipac.caltech.edu} we further verified the cluster redshifts using the galaxy redshifts. Galaxy redshifts were compiled from: NED \citep[mostly from the 6dF survey, ][]{6dF}; the 2MASS redshift survey \citep{TMZ}, the WINGS survey \citep{WINGSdata}; the CAIRNS survey \citep{Rines2003}. We found that the NED redshifts for the clusters Abell~838, Abell~1631 and DC2349-28 (0.0502, 0.01394 \& 0.0648, respectively) were inconsistent with the redshifts of all the galaxies. A detailed investigation of the redshift distributions for each revealed that they are each most likely projections of two or more clusters along the line of sight.

For Abell~838, only 19 of the 62 galaxies were successfully cross-matched with the aforementioned redshift surveys. For these 19 galaxies, the redshift distribution is bimodal and equally distributed at z$\approx$0.023 \& z$\approx$0.052. However, when considering all galaxies in the spectroscopic catalogues within a radius of 0.8\degree, more than twice as many galaxies are associated with the z$\approx$0.023 over-density than the one at z$\approx$0.052. As such, we associate Abell~838 with a cluster at z=0.023. For Abell~1631, we successfully cross-match 124 of the 139 galaxies in D80 and find that only 17 are associated with an over-density at z=0.014, while 103 are associated with an over-density at z$\approx$0.047. Accordingly, we associate Abell~1631 with a cluster at z=0.047. Finally, for DC2349-28, we cross-match 55 of 68 galaxies and find 28 galaxies associated with an over-density at z$\approx$0.03, and 22 galaxies associated with a number of smaller over-densities at 0.045$<$z$<$0.075. When considering all galaxies with redshifts within a radius of 0.93 \degree, we find that twice as many are associated with the z$\approx$0.03 over-density compared to the 0.45$<$z$<$0.75 structures. 

\bsp

\label{lastpage}

\end{document}